# TOFHIR2: The readout ASIC of the CMS Barrel MIP Timing Detector

E. Albuquerque[e], M. Araújo[d], A. Benaglia[c], A. Boletti[d], R. Bugalho[e], T. Coutinho[e], F. De Guio[c,f], P. Faccioli[d], L. Ferramacho[e], M. Firlej[a], T. Fiutowski[a], R. Francisco[e], M. Gallinaro[d], A. Ghezzi[c,f], J. Hollar[d], M. Idzik[a], H. Legoinha[d], N. Leonardo[d], C. Leong[e], M.T. Lucchini[c], M. Malberti[c], G. Marozzo[d], G. Da Molin[d], J. Moron[a], T. Niknejad[d,e], L. Oliveira[b], N. Oliveira[e], S. Palluotto[c,f], M. Pisano[d], N. Redaelli[c], C. Silva[d], J. C. Silva[d,e], R. Silva[d,e], M. Silveira[e], K. Swientek[a], T. Tabarelli de Fatis[c,f], S. Tavernier[e], J. Varela[d,e], V. Varela[e], S. White[g], J. Wulff[d]

[a] *AGH University of Krakow, Faculty of Physics and Applied Computer Science, Krakow, Poland*
[b] *DEE, CTS-UNINOVA, FCT-UNL, Caparica, Portugal*
[c] *INFN Sez. Milano Bicocca, Italy*
[d] *LIP, Lisbon, Portugal*
[e] *PETsys Electronics, Oeiras, Portugal*
[f] *University of Milano-Bicocca, Milano, Italy*
[g] *University of Virginia, Charlottesville, Virginia, USA*

E-mail: joao.varela@cern.ch

ABSTRACT: The CMS detector will be upgraded for the HL-LHC to include a MIP Timing Detector (MTD). The MTD will consist of barrel and endcap timing layers, BTL and ETL respectively, providing precision timing of charged particles. The BTL sensors are based on LYSO:Ce scintillation crystals coupled to SiPMs with TOFHIR2 ASICs for the front-end readout. A resolution of 30-60 ps for MIP signals at a rate of 2.5 Mhit/s per channel is expected along the HL-LHC lifetime. We present an overview of the TOFHIR2 requirements and design, simulation results and measurements with TOFHIR2 ASICs. The measurements of TOFHIR2 associated to sensor modules were performed in different test setups using internal test pulses or blue and UV laser pulses emulating the signals expected in the experiment. The measurements show a time resolution of 24 ps initially during Beginning of Operation (BoO) and 58 ps at End of Operation (EoO) conditions, matching well the BTL requirements. We also showed that the time resolution is stable up to the highest expected MIP rate. Extensive radiation tests were performed, both with x-rays and heavy ions, showing that TOFHIR2 is not affected by the radiation environment during the experiment lifetime.

**Contents**

## 1. Introduction

The Upgrade of the CMS experiment [1] for the future High-Luminosity phase of the Large Hadron Collider at CERN (HL-LHC) includes the construction of a new MIP[1]Timing Detector (MTD) to measure the time of charged particles with high precision [2]. The MTD barrel section (Barrel Timing Layer - BTL) is a thin cylindrical detector with a total active surface of about 38 m$^2$ based on LYSO:Ce scintillating crystals [3] coupled to silicon photomultipliers (SiPM) [4]. The individual cell consists of a crystal bar with approximately 3.0×3.75×54.7 mm$^3$ with two ~3.0×3.0 mm$^2$ SiPMs glued at each end. The SiPM signals are conditioned and digitized by the TOFHIR2 integrated circuit described in this paper. The full BTL detector has 331'776 electronics channels.

LYSO is a scintillating crystal with high density (7.1g/cm$^3$), large light yield (~40 thousand photons/MeV) and scintillation decay time of about 40 ns. A charged particle crossing the detector deposits in the crystal on average an energy of 4.2 MeV[2]. A fraction of the scintillation light is detected at the two ends of the crystal bar by SiPMs. The individual SiPM photo-sensor has 14'400 micro-cells of 25x25 μm$^2$ each. The Photon Detection Efficiency (PDE) at 420 nm is 20-55% and the gain is 2.9-9.7×10$^5$ for an overvoltage[3] in the range 0.9-3.1V. The SiPMs are operated with overvoltage in this range optimizing PDE and gain versus Dark Count Rate (DCR). The pulse amplitude is proportional to the number of photons emitted by the particle interacting with the crystal. The current pulse from the SiPM is a unipolar signal with peaking time of 30 ns and tail of approximately 200 ns.

---
[1] MIP - Minimum Ionizing Particle

During its operation, the detector is exposed to a large flux of particles (reaching an integrated fluence of $2\times10^{14}$ n$_{eq}$/cm$^2$ by end of operation), which creates defects in the SiPM silicon crystal. These defects are responsible for an increase of DCR from 0.5 MHz before irradiation up to 10-20 GHz at the end of the detector operation. To reduce the DarkCount noise the SiPMs are operated at a temperature of -45°C and will be periodically annealed at 60°C.

The main requirements to the TOFHIR2 ASIC are: (1) to measure MIP timing with a precision of the order of 30 (60) ps at the beginning (end) of HL-LHC operation; (2) to mitigate the effect of large DCR implementing noise suppression circuitry; and (3) to provide a measurement of the signal amplitude with <5% precision for time-walk corrections. The chip must cope with a MIP input rate of 2.5 Mhit/s per channel and a rate of low energy hits (<1 MeV) up to 5 Mhits/s per channel. Additionally, it should be able to accept the CMS Level-1 trigger and synchronization signals, with output bandwidth of 640 Mb/s. The chip should have power consumption lower than 20 mW per channel. Furthermore, the TOFHIR2 must be resistant to the expected total ionization dose TID (2.9 Mrad) and to the integrated particle fluence ($2\times10^{14}$ n$_{eq}$/cm$^2$) in the BTL. A summary of the specification parameters is given in Table I.

**Table I** - TOFHIR2 specification parameters.

| | |
|---|---|
| Number of channels | 32 |
| Voltage supply | 1.2 V |
| Reference voltages | Internal |
| Radiation tolerance | Yes |
| DarkCount noise filter | Yes |
| Number of TAC and QAC | 8 |
| TDC bin (ps) | 10 |
| 10-bit SAR ADC (MHz) | 40 |
| Max MIP rate/ch (MHz) | 2.5 |
| Max low E rate/ch (MHz) | 5 |
| Output bandwidth (Mb/s) | 640 |
| Power consumption/channel (mW) | <20 |
| Tolerance to TID (Mrad) | >3 |
| Tolerance to particle fluence (n$_{eq}$/cm$^2$) | >2×10$^{14}$ |

TOFHIR2 is a new chip designed by PETsys Electronics [5][4] in CMOS 130 nm technology and fabricated in the same foundry where the HEP community is focusing the production of the ASICs for the HL-LHC upgrades. TOFHIR2 follows a first implementation with limited functionality (TOFHIR1) adapted from an existing chip developed by PETsys Electronics for PET applications in technology CMOS 110 nm technology of a different foundry [6]. The CMOS 130 nm technology was adopted for the final chip given its good behavior under irradiation (small shifts of the transistors threshold voltage and low leakage current). The first full version (TOFHIR2A) was developed and tested in 2020 [7], followed by TOFHIR2X in 2021 which implemented a full current-mode, noise suppression analog front-end [8]. The last prototype (TOFHIR2B), implementing improved protection against Single Event Effects (SEE) induced by radiation, was developed before the final production version (TOFHIR2C) in 2023. In this paper, we present a summary of the design principles of the TOFHIR2 ASIC and the results of the measurements performed[5]. Results obtained in test beam experiments with final sensor modules and TOFHIR2C shall be reported in another paper.

---

[2] Most Probable Value of the energy deposition averaged over the entire BTL.
[3] Overvoltage is the difference between the SiPM operation voltage and the breakdown voltage.
[4] The design of the 10-bit SAR ADC was provided by AGH, Krakow.

## 2. State of the art

The sensor technology based on LYSO crystals and SiPMs adopted for BTL has been used during the last twenty years in medical imaging Positron Emission Tomography (PET). In PET, radioactive tracer molecules are injected into the patient's body. The decay of tracer molecules generates a positron-electron annihilation event resulting in two back-to-back γ-photons[6] with an energy of 511 keV that are absorbed by a cylindrical array of LYSO crystals coupled to SiPMs surrounding the patient [9][10]. The LYSO crystal converts the incident particle into optical photons whose number is proportional to the energy deposited in the crystal. A fraction of the photons is detected at the SiPM. Detected photons are then converted to electrons[7] that are multiplied in avalanche by the SiPM, generating a current pulse that can be discriminated and digitized to obtain a measurement of the time of the particle detection, referred to as the "timestamp". The SiPMs used in PET applications (typically with micro-cells of 50x50 μm$^2$) have a gain ten times larger than the SiPMs optimized for use under extreme radiation in BTL, where the maximum allowed SiPM gain is determined by the allowed SiPM power consumption due to DCR.

The timestamp of the detected interaction is determined from the first optical photons detected by the SiPM[8]. PET systems can measure the difference in arrival times of two 511 keV gamma rays with a resolution the order of few hundred picoseconds (FWHM) [11]. This allows an improvement of the signal-to-noise ratio (SNR) of the image [12][13]. The precise digitization of event timestamps and γ-photon energies is typically achieved by employing application-specific integrated circuits (ASICs) [14]-[30]. Table II shows a comparison of the main characteristics of several ASICs developed for PET applications. However, none of these circuits is tolerant to radiation.

---

[5] The results presented in this paper have been obtained mostly with the final version TOFHIR2C. When noted, we present results obtained with the previous versions of the ASIC.
[6] γ-photons or gamma-rays have energy above the x-ray band
[7] Usually referred to as photoelectrons (p.e.)
[8] Due to the long decay time of the LYSO scintillating light only the first optical photons are relevant for a time measurement with resolution three orders of magnitude smaller than the decay time.

**Table II** – Comparison of ASICs for PET applications (adapted from [19])

| ASIC | Timestamp digitization | Power (mW/ch) | CTR FWHM (ps) | Crystal height (mm) | Ref. |
|---|---|---|---|---|---|
| FlexToT | external | 11 | 123 | 5 | [27][28] |
| HRFlexToT | external | 3.5 | 180 | 20 | [29] |
| STiC3 | TDC on ASIC | 25 | 240 | 15 | [21] |
| PETA4 | TDC on ASIC | <40 | 460 | 25 | [26] |
| Petiroc | external | 3.5 | n.a. | n.a. | [22][22] |
| Triroc | TDC on ASIC | 10 | 432.7 | 10 | [23][24] |
| TOFPET1 | TDC on ASIC | 8-11 | 290.7 | 15 | [15][16] |
| TOFPET2 | TDC on ASIC | 3.6-7.2 | 119 | 3 | [6][17][18][19] |

Along the detection chain several effects introduce stochastic fluctuations that determine the detector time resolution, namely the photo-statistics, the SiPM DCR, the analog front-end (AFE) noise and the time digitization. The contribution from photo-statistics is related to the stochastic fluctuation in the time-of-arrival of photons detected at the SiPM and scales with $1/\sqrt{N_{pe}}$, where $N_{pe}$ is the total number of detected photoelectrons. The contribution to the time resolution of DarkCount noise scales with $\sigma_{DCR}/N_{pe}$, where $\sigma_{DCR}$ is the standard deviation of the DarkCount noise. $\sigma_{DCR}$ scales with $\sqrt{DCR}$. The contribution of the front-end noise is given by $\sigma_{AFE}/SR$, where $\sigma_{AFE}$ is the standard deviation of the front-end noise and SR is the slew rate of the pulse rising edge at the level of the discrimination threshold. These contributions are not correlated and therefore the total time resolution $\sigma_t$ is obtained by the quadratic sum of them:

$$\sigma_t = \sigma_t^{phot} \oplus \sigma_t^{DCR} \oplus \sigma_t^{AFE}$$

The contribution arising from the time-to-digital conversion (including the clock jitter) is normally of the same order of the front-end noise or smaller. The jitter of the SiPM response is small when compared to the other sources of time uncertainty.

In bench-top experiments using optimized discrete circuits associated to small LYSO crystals and SiPMs with improved single-photon time resolution (SPTR) it was possible to measure a Coincidence Time Resolution (CTR) of the two PET γ-photons of the order of 60 ps FWHM, which corresponds to a time resolution per channel of about 20 ps r.m.s (standard deviation)[9] largely dominated by photo-statistics [31]. In measurements performed with PET ASICs, typically the best CTR is of the order of 120 ps FWHM (see for example [6]), corresponding to single channel resolution of 36 ps r.m.s, which results from the contributions of photo-statistics, front-end noise and digitization, estimated at around 20 ps each, added quadratically.

In BTL before irradiation, the target contributions to the time resolution are 20 ps from photostatistics, 10 ps from front-end noise and 10 ps from the digitization, for a total of 25 ps [2]. After accounting for the various factors contributing to the pulse amplitude (energy deposit, light collection efficiency, SiPM PDE and Gain), pulses in BTL before irradiation are about five times smaller than in PET. Therefore, achieving a noise contribution $\sigma_{AFE}/SR$ two times smaller than in PET applications requires one order of magnitude improvement of the AFE by reducing the solid-state noise and increasing the bandwidth.

---

[9] In PET applications, the time resolution is usually given by the full width at half maximum (FWHM) of the distribution of the time difference of the two γ-photons (CTR). The conversion factor between FWHM and root mean square (RMS) for a Gaussian distribution is 2.35. The single channel resolution and the CTR differ by a factor $\sqrt{2}$.

During BTL operation the DarkCount noise increases by several orders of magnitude due to radiation, as already mentioned. The time resolution is progressively dominated by the term $\sigma_t^{DCR}$. To achieve the goals of the BTL detector, it is necessary to implement noise suppression in the AFE to mitigate the negative impact of the DCR, by a factor larger than 2. The AFE of TOFHIR2 implements DarkCount noise suppression, for the first time in radiation detectors. On the other hand, in order to limit the increase of SiPM current due to DCR, staying within the power limits of the SiPM bias voltage supply system, the SiPM overvoltage is decreased from about 3V to 1V during the detector operation. The consequent drop of SiPM PDE and gain translates into a decrease of the number of photoelectrons by a factor larger than two and of pulse amplitude by a factor larger than five, which is responsible for the deterioration of $\sigma_t^{phot}$ and $\sigma_t^{AFE}$.

Accumulated exposure of silicon chips to ionizing radiation creates trapped charges at oxide/silicon interfaces causing quasi-permanent device shifts that may lead to chip malfunction [32]. The degradation depends on the device technology, process, and bias conditions. The effects due to TID in the technology used in TOFHIR2 have been characterized in [33][34]. The PMOS devices lose their current drive capability by up to 5% at 1 Mrad and 10% at 10 Mrad. NMOS devices have increased channel leakage of up to four orders of magnitude in the 1~10 Mrad range. NMOS and PMOS also experience threshold voltage shifts. These effects are higher in the 1~10 Mrad range. They depend strongly on the device size but also on the site where the chips are fabricated. Versions of TOFHIR2 have been fabricated in two different foundries of the same company, named in the following as A and B. Foundry A has considerably lower TID sensitivity than foundry B [33].

Single event upsets and transients, usually referred to as single event effects (SEE), caused by local ionization created by recoiling ions hit by heavy particles (mainly protons and neutrons) can cause a chip either to "lock-up", to lose synchronization or to modify the configuration, thus requiring a time-consuming reset process. These effects are strongly dependent on the ion charge and therefore, induced ionization charge. The techniques used in TOFHIR2 to mitigate the radiation effects are described in the next section.

## 3. Architecture and implementation

The functional block diagram of the TOFHIR2 ASIC is shown in Figure 1. The ASIC has 32 independent channels, a service block, and global control and trigger logic. Each channel integrates the analog front-end, time and charge digitizers and the channel digital control. The service block provides bias currents, several reference voltages and currents, and monitoring features. The global control and trigger logic handles the chip configuration, the digital data flow and transmission, and the trigger filtering (CMS Level 1 trigger).

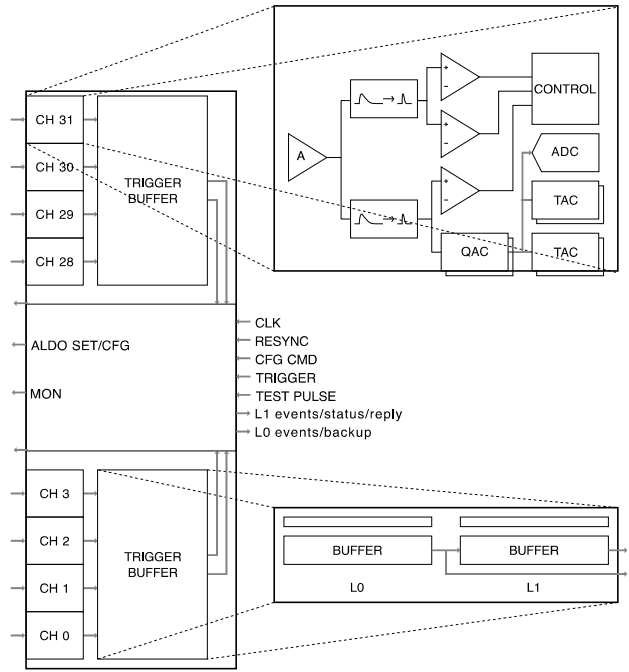

**Figure 1.** Functional block diagram of the TOFHIR2 ASIC

## 3.1 Analog design

A block diagram of one of the 32 TOFHIR2 channels is shown in Figure 2. Each ASIC channel contains one pre-amplifier, two post-amplifiers (T and E), three leading edge discriminators (T1, T2 and E), two Time-to-Amplitude Converters (TAC), one Charge-to-Amplitude Converter (QAC), one 40 MHz 10-bit SAR ADC and local control logic. TACs and QACs are replicated eight-fold to handle Poisson fluctuations of the event[10] rate. The pre-amplifier provides a low impedance input to the sensor's current signal. The input current is replicated into three branches for timing, energy discrimination and charge integration. Pulse filtering is included in the post-amplifiers to mitigate the deterioration of time resolution due to the large DCR induced by radiation and due to pile-up of LYSO pulse tails, as described below. For testing purposes, the chip implements an analog test pulse generator.

---

[10] Unless otherwise specified, in this paper the term event refers to a signal at the channel input.

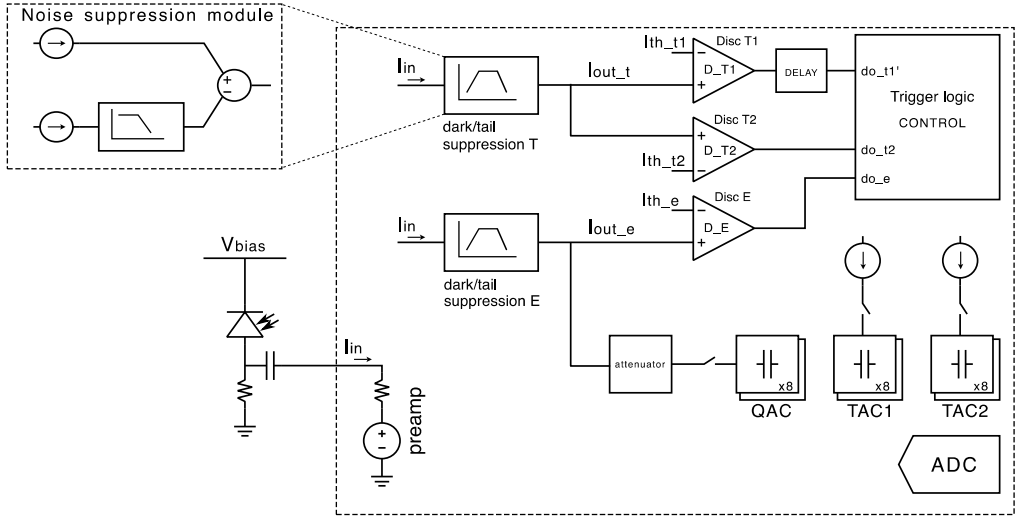

**Figure 2.** Block diagram of the TOFHIR2 channel.

To obtain the required time resolution, since the SiPMs output is a current, the developed circuit is designed to process the input current without the usual conversion to voltage. Full current-mode increases the bandwidth, since the signal flows through low impedance nodes, making the circuits less sensitive to parameter variations and parasitic elements, thus reducing the mismatch between channels.

The preamplifier is a current buffer to isolate the large SiPM load from the core of the AFE. The output current is mirrored to different timing and energy branches. Considering the effect of SiPM DCR, time-tagging relevant events directly at the preamplifier output with a time resolution better that 60 ps at EoO is impossible without proper signal processing. The challenge arises from the fact that signal events and DarkCount pulses have the same shape and frequency content, although they differ in amplitude.

In addition, the high event rate hitting the preamplifier leads to baseline drifts due to severe piling-up of residual pulse tails. Since event flagging is done with comparators that have their threshold set to a fixed value referenced to the preamplifier baseline, baseline drifts will lead to wrong comparator firing time. Therefore, the baseline must be very stable for accurate timing.

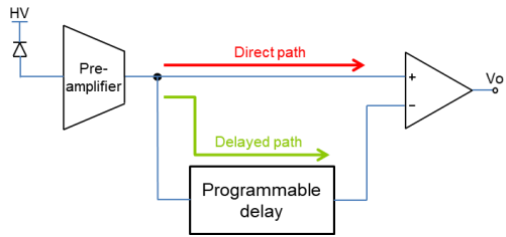

**Figure 3.** DLED block diagram.

Motivated by the above considerations, we used the signal processing technique DLED (Differential Leading Edge Discriminator) [35], implemented here for the first time in a CMOS integrated circuit (Figure 3 and Figure 4). With appropriate delay settings, DLED preserves the rising edge of the pulse and cancels the pulse tail (Figure 5). This approach reduces DarkCount noise and avoids baseline fluctuations and pile-up, since the input unipolar signal is converted to a bipolar signal. As shown in Figure 4, at the DLED output node Io is the current difference (I1-I2) between the direct and delayed paths, hence the DLED signal processing in full current mode.

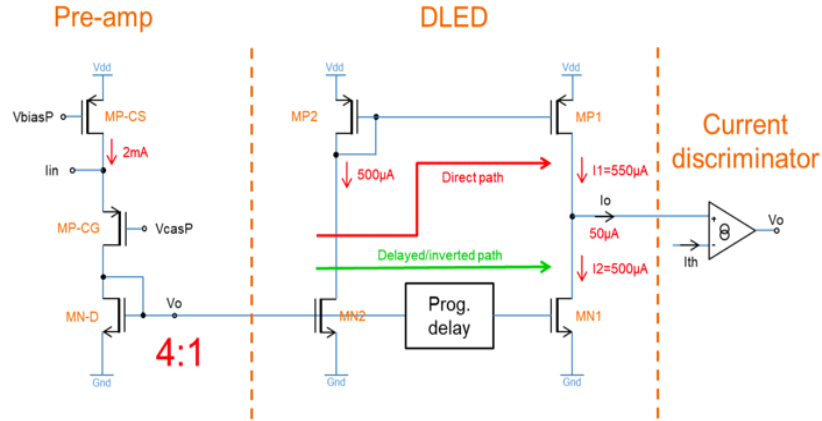

**Figure 4.** Full current mode AFE.

An ideal delay line cannot be implemented in a CMOS process; hence we have used a series of RC taps behaving as a low pass filter. In TOFHIR2 a delay line with eight taps can be programmed to have an equivalent delay in the interval between 200 and 1400 ps. The ability to vary the delay value also provides an additional degree of freedom to fine tune the pulse shape, since increasing the delay increases the pulse amplitude.

The pulse shape also depends on the fabrication process mismatch between PMOS device MP1 and NMOS device MN1 (Figure 4). A pulse calibration feature has been implemented via an array of PMOS devices in parallel with MP1 forming a 5-bit DAC, which can be programmed to trim the pulse shape. Since the BTL detector will have more than 300k channels, it is important to have a quick and expeditious trimming process to set the pulse shape on demand. This can be done by simply measuring the width of the pulse lobe above the baseline using the T1 comparator and the two TDCs.

At the DLED output, the baseline is stabilized with a baseline holding feedback loop. The baseline holder amplifier has a single low frequency pole in the mHz region, therefore the dynamics of the baseline holder does not modify the pulse shape.

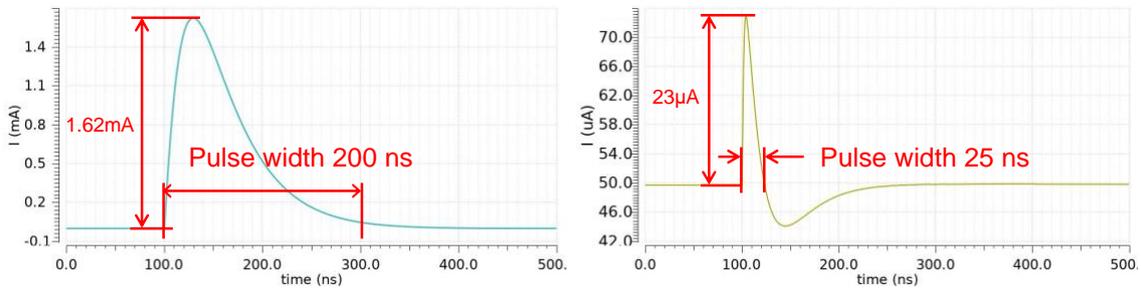

**Figure 5.** DLED input (left) and output (right) waveform. The example is for EoO conditions.

Table III shows the results of a simulation in one possible scenario for the end of operation (EoO) conditions (DCR 55 GHz, signal yield of MIP pulse 6000 p.e., SiPM gain $1.5 \times 10^5$), comparing the time resolution, estimated by $\sigma_{DCR}$/SR at the optimal threshold, at the output of the SiPM and at the output of the DLED module. An improvement by a factor 3.5 is obtained.

**Table III** – Simulation of the time resolution of SIPM (DLED input) and DLED output current pulses.

|  | SiPM output | DLED output |
|---|---|---|
| Slew rate ($\mu$A/ns) | 135.9 | 9.93 |
| Noise r.m.s ($\mu$A) | 24.5 | 0.51 |
| $\sigma_{noise}$/ SR (ps) | 180 | 52 |

In order to trigger the timing information retrieval process of an incoming pulse, the output current of the DLED is mirrored into two comparators T1 and T2 (Figure 2). T1 is set to a low threshold where SR is the highest to maximize the timing resolution, and T2 is set to a higher threshold value for low energy event filtering. To cope with the large DCR and low energy hits, the output of T1 is delayed to ensure that the timing measurement is triggered only for pulses above the T2 threshold. The current comparator from [36] with positive feedback and a low impedance output node is used.

The T1 and T2 threshold current DACs have four configurable LSB settings [0.156 µA, 0.313 µA, 0.47 µA, 0.63 µA] and [0.313 µA, 0.63 µA, 0.94 µA, 1.25 µA], respectively. All the DAC transfer curves start below the baseline at 45µA so that any baseline spread that may occur is within the DAC range. The lowest T1 threshold range is [45,55]µA and the corresponding LSB value is 0.156µA, and it is mainly used for noise measurement by scanning the DAC across the baseline and recording the number of hits (S-curve).

The common-gate preamplifier current is mirrored to two DLED blocks: one used for timing measurements and the other used for energy measurements. The energy branch (E) is not as demanding in terms of speed as the timing branch (T), however it must have a good linearity. Considering that the current pulse from the SiPM has a large dynamic range, the output current of the energy branch DLED is mirrored into a programmable gain stage with eight settings (2/8, 3/8, .., 9/8). This gain stage is needed so that the pulses can be properly framed within the dynamic range of the integrator, maximizing energy resolution. The output current of the energy branch DLED is also mirrored into a comparator E for pulse energy discrimination. The threshold DAC for energy discrimination is the same as T2. A more complete description of the TOFHIR2 analog front-end can be found in [37].

The active integration block (QAC) integrates the current exiting the energy branch DLED in a configurable integration window that typically has a value of 25 ns (corresponding to the LHC bunch spacing). The dynamic range of the integrator is 400 mV. The current has two components: the baseline component (1) upon which the pulse (2) is superimposed. It is mandatory to have a baseline cancellation that removes the DC value to maximize the dynamic range of pulse integration. The current being subtracted is supplied by a current DAC with a LSB of 80 nA. Hence the residue will always be lower than 80 nA and the maximum pedestal for a 25 ns integration window is just 4 mV.

The TDC is composed of a time-to-amplitude converter (TAC) followed by the analog-to-digital converter (ADC). Based on the discriminator outputs, the control logic controls the integration of a constant reference current in the time to amplitude converter (TAC) which provides the fine time measurement. On the rising edge of a trigger signal a switch is closed, charging an analog buffer (as shown in Figure 2). On the next rising edge of the clock the switch opens, stopping the charging process. The value of a global clock counter is latched on the next clock providing value $t_{coarse}$. If the event is valid ($I_{out\_E} > I_{th\_E}$), the voltage stored in the TAC will be digitized as $t_{fine}$. The two TDCs operate identically, using any of the leading or falling edges of the trigger signals chosen by configuration. The design TDC time binning is 10 ps.

Each channel has 8 sets of TAC and QAC analog buffers where analog values are stored before being digitized. Every time the channel goes through "rearm" state, a new buffer set is selected round-robin. This allows the channel to rearm without waiting for the previous event to be digitized. When an event is valid and digitized, the number of the analog buffer set used to process that event is transmitted along the event.

The content of the analog buffers shifts slowly due to leakage effects, which affects the time and charge measurements. To keep this effect under 0.1 LSB, whenever the channel has been in ready state for 100 µs, an invalid event is triggered causing the channel to rearm with the next, fresh, set of buffers.

A SAR ADC with 10-bit and sampling frequency of 40 MHz was implemented in TOFHIR2. The design is based on a fully differential capacitive DAC as proposed in [38] providing high noise rejection, twice the dynamic range (2×1.2 V) and low power consumption. In TOFHIR2, the SAR ADC is operated in asynchronous mode performing the conversions when requested by the digital logic.

Additionally, TOFHIR2 implements two 8-bit DACs to adjust the two SiPM bias voltages provided by the ALDO2 ASIC [45]. The DACs can be configured in two ranges, high range (0.74 V-0.98 V) and low range (0.82 V-0.94 V) for finer bias voltage adjustment. The ALDO2 ASIC multiplies these voltages by a factor of the order of 40 such that the LSB of the SiPM bias adjustment can be as low as 20 mV. The spread of the DAC voltages in corner simulations is below 0.6%. The voltage spread in a temperature sweep [-40,+100] ºC (typical process corner and $V_{DD}$ = 1.2 V) is of the order of 0.3%, that can be translated to 20 µV/ºC. TOFHIR2 provides two control bits to ALDO2, to enable the bias voltages and to select the range in the ALDO2 bias current monitoring.

In the development of TOFHIR2 we used several mitigation techniques to minimize the impact of TID radiation. Following good practice, we used transistor gate lengths greater than 3×$L_{min}$, where $L_{min}$ is the minimum gate length of the technology. Where maximum performance and bandwidth was required (signal path) and if L < 3×$L_{min}$ was necessary, Enclosed Layout Transistor (ELT) were used for additional radiation hardness [39][40][41].

### 3.2 Digital design

In each channel, the output of the three discriminators connects to the channel trigger generator which generates the trigger signals controlling the TDCs and QDC digitization and the data transfer to the output FIFOs (Figure 2). TOFHIR2 implements a multi-level event trigger and rejection scheme. In the nominal operation mode:
- Event time is measured at a low threshold (T1)
- Events which do not trigger threshold T2 are rejected without any dead time.
- Events which trigger T2 but not E are rejected with less than 25 ns dead time.
- Only events that trigger all three thresholds are considered valid and digitized.

TOFHIR2 implements in each channel a 24-bit event counter that can be configured to perform several functions using the output of the discriminators, namely discriminator calibration and noise measurement, low energy event monitoring, and event loss monitoring. When the counters are activated by configuration the content of the counters are transmitted periodically as a special data packet.

The TOFHIR 2 digital interface comprises the following digital signals:
1. Clock and fast signals: Clock 160 MHz; Resync; Trigger (L0 and L1); Test pulse.
2. Configuration: 80 Mbit/s multi-drop link.
3. Output: primary link (320 Mbit/s); secondary/backup link (320 Mbit/s).
4. Static configuration pins: chip ID; RX alignment mode.

The Resync signal serves three purposes depending on its length, namely reset the time-tag counter only, reset the time-tag counter and clear the event processing chain, and full reset including the chip configuration. The system level synchronization is based on the TOFHIR2 coarse time-tag, which counts clock cycles (160 MHz) since the last Resync signal. TOFHIR2 has no notion of CMS event[11]/orbit/bunch number. The backend should map between the TOFHIR2 coarse time-tag and the CMS event identifiers. Since the backend generates the TOFHIR2 Resync, the backend can keep a copy of the TOFHIR2 time-tag counter and its mapping to CMS event identifiers (Figure 6).

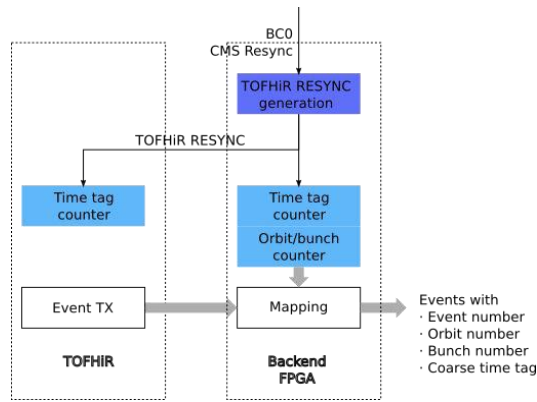

**Figure 6.** Synchronization of TOFHIR2 data.

The trigger requests have the form of a simple yes/no bitstream at 80 Mb/s. In each 25 ns bin, two bits define the Level 0 and the Level 1 trigger status. The Level 0 pre-trigger is intended to capture BTL data to be used in the Level 1 trigger algorithms. The latency of the L0 and L1 triggers is configurable.

Each TX link has a bandwidth of 320 Mbit/s transmitting in Double Date Rate mode. The data is 8B/10B encoded. The use of the two links is configurable. The primary link can be used to transmit L1 filtered data, while the second link can transmit L0 filtered data. But other options are possible, for example, to transmit L1 data through the second link in case of malfunction of the primary link, or to allocate both links to the transmission of L1 data if extra bandwidth is required.

The chip has an input for an external digital test pulse. It can either receive a test pulse generated by an external system or it can receive a test pulse clock which drives an internal test pulse generator with configurable period and duration. The phase of the internally generated test pulse can be shifted by shifting the phase of the test pulse clock relatively to the main chip clock[12]. The internal test pulse can be configured to trigger the logic controlling the TDCs and QDC or the analog pulse generator into pre-amplifier.

---

[11] In this context, event refers to a LHC bunch crossing flagged by the CMS L1 Trigger.

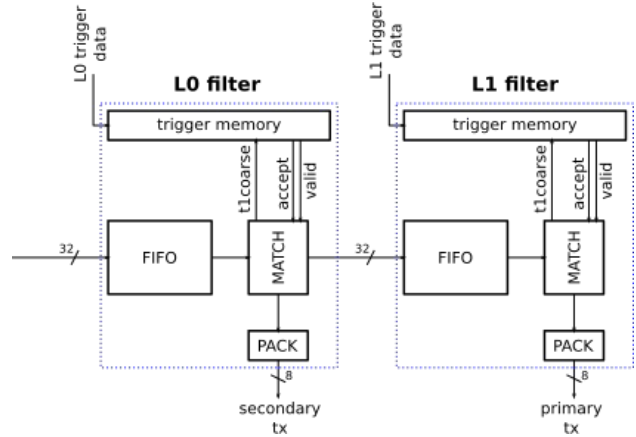

**Figure 7.** L0 and L1 trigger event filtering.

TOFHIR2 implements event filtering based on the external L0 and L1 trigger inputs. Its functioning is illustrated in Figure 7. Events are stored in a FIFO and the L0/L1 trigger bits are stored in a circular memory. The matching logic looks-up for the trigger bits in the trigger memory addressed by the event coarse time-tag. If the event is not wanted, the event data is moved forward to the next trigger only. If the event is wanted, the event data is moved forward to both the next level and the transmission link. The implementation of the two-level triggering implied small overhead since the L1 latency defines the total buffer requirements.

The chip configuration uses the 80 Mb/s RX link. The commands are 8B/10B encoded. Up to 32 chips can share (multi-drop) the same configuration link making use of the 5-bit chip ID. In total there are 35 configuration registers, one per channel and three global configuration registers. The transmission of replies to commands is multiplexed into the primary TX path.

To eliminate SEEs, we implemented Triple Modular Redundancy (TMR) in all configuration logic and state machines [42]. To protect against the effect of single event transients, we also added TMR in the clock and resync distribution tree. Given the very low probability of data corruption, the data path was not protected to save silicon area. TMR was implemented using the CERN TMRG tool [43]. TMR correction could fail if a highly ionizing particle flips the state of two of the three TMR flip-flops, or in case the (asynchronous) error detection is not synchronized to the correction clock edge leading to metastability. In TOFHIR2 a chain of four registers is used to synchronize the asynchronous detection to the correction clock.

## 4. Simulation results

The expected time resolution is obtained from a detailed simulation of the ASIC associated to the sensor. The ASIC inputs are AC connected to the equivalent electrical model of the SiPM depicted in Figure 8. Early simulation results were obtained for SiPM with microcells of

---

[12] In the BTL electronics system, the clocks are distributed by the lpBGT chip which can provide phase adjusted clocks.

15×15 μm². The most recent results were obtained with microcells of 25×25 μm², the final option for the BTL detector.

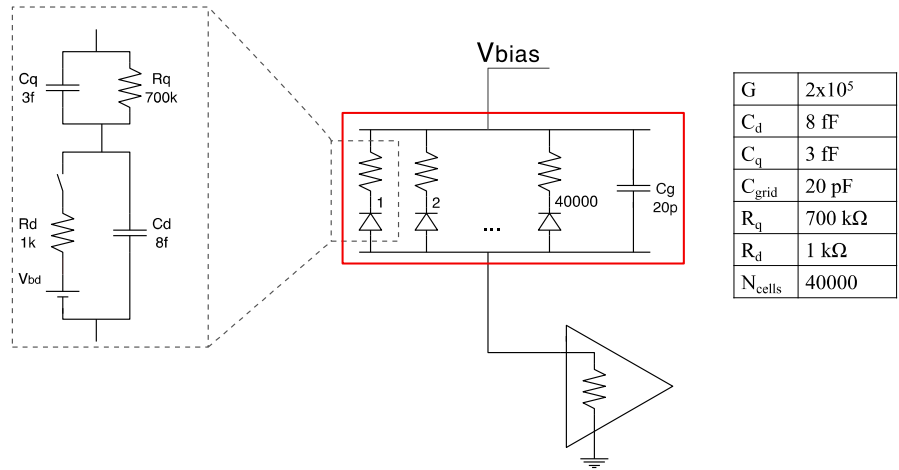

**Figure 8.** SiPM equivalent electrical model assumed in the TOFHIR2 simulations. The values indicated in the table (right) correspond to SiPM microcells of 15×15 μm².

Several points along the detector lifetime are simulated reproducing the evolution of pulse amplitude and DCR due to radiation. During the TOFHIR2 design phase, carried-on in parallel with the sensor development, we have assumed that: at BoO, the MIP pulses yield 9500 photoelectrons on average, the SiPM gain is $3.8\times10^5$ and the DCR is negligible; at EoO, after the detector has been exposed to a fluence of $2\times10^{14}$ $n_{eq}\cdot cm^{-2}$, the MIP pulses yield 6000 photoelectrons, the DCR is of the order of 55 GHz, and SiPM gain is $1.5\times10^5$. The SiPM pulse simulation was performed assuming 15×15 μm² microcells.

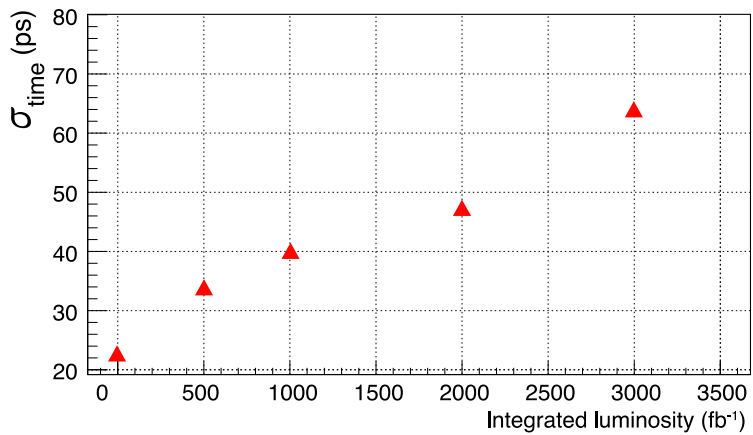

**Figure 9.** Simulation of time resolution throughout the HL-LHC operation. Due to radiation, the SiPM DCR increases by more than four orders of magnitude at end of operation. To control this increase, the SiPM overvoltage is reduced gradually, which results in a decrease of Photon Detection Efficiency (PDE) and consequently in a decrease of signal yield.

Figure 9 shows the expected MIP time resolution (for double readout of a LYSO bar) along the detector operation obtained with a full TOFHIR2 simulation that includes the effects of photo-statistics, SiPM electrical model and jitter, electronics noise, DCR and TDC binning. The data points in Figure 9 correspond to the conditions specified in Table IV. We concluded that under the assumed conditions the time resolution is expected to be of the order of 25 ps at BoO degrading to about 65 ps at EoO, confirming the expectations described in the BTL Technical Design Report (TDR) [2].

**Table IV.** Conditions used in the simulation of the data points in Figure 9.

| Integrated Luminosity (fb$^{-1}$) | Number of p.e. | SiPM gain | DCR (GHz) |
|---|---|---|---|
| 0 | 9500 | $3.8 \times 10^5$ | 0 |
| 500 | 9000 | $2.9 \times 10^5$ | 20 |
| 1000 | 8000 | $2.5 \times 10^5$ | 30 |
| 2000 | 7000 | $1.9 \times 10^5$ | 45 |
| 3000 | 6000 | $1.5 \times 10^5$ | 55 |

Along the sensor development process, it was realized that the photoelectron yield was about 50% of the estimations described in the TDR, thus significantly deteriorating the time resolution, in particular at the EoO. To cope with this fact, an intense R&D program was carried out in 2022 which led to several improvements of the sensor module. Several modifications allowed the expected timing performance to be recovered, namely:
- larger thickness of the LYSO crystal bars to increase the MIP energy deposit;
- SiPMs with larger microcells (25 μm) allowing an increase of the SiPM gain and consequently the slew rate of the pulse rising edge;
- using silicon-based Themo-Electric Coolers (TEC) mounted on the back of the SiPM array to reduce the SiPM operating temperature from -35°C to -45°C, which reduces the DCR by about a factor of two [46];
- improving the S/N ratio of the TOFHIR2 electronics, by modifying the current mirroring between the preamp and the DLED block (implemented in TOFHIR2B and in the final version TOFHIR2C).

Several simulations of TOFHIR2X, TOFHIR2B and TOFHIR2C coupled to the new sensor module have been performed during these developments confirming the options taken and supporting the measurements described in section 6.

**5. Experimental setup**

The TOFHIR2 ASICs are assembled in the BTL front-end board (Figure 10). The chip dimensions are 8.5×5.4 mm$^2$, the total number of pads is 258, the channel pitch is 225 μm and the pad pitch is 100 μm. The chip is assembled in a Ball Grid Array (BGA) package of 14×14 mm$^2$. Each board includes two TOFHIR2 ASICs, two ALDO2 ASICs (implementing LV and BV regulators) and four SiPM-array input connectors. Each sensor module (Figure 11) is composed of 16 LYSO crystal bars glued at both ends to linear arrays of 16 SiPMs. The sensor module used in these measurements has a slit on the external wrapping to allow the excitation of the crystal scintillation light with a UV laser (375 nm) emulating the MIP energy deposit.

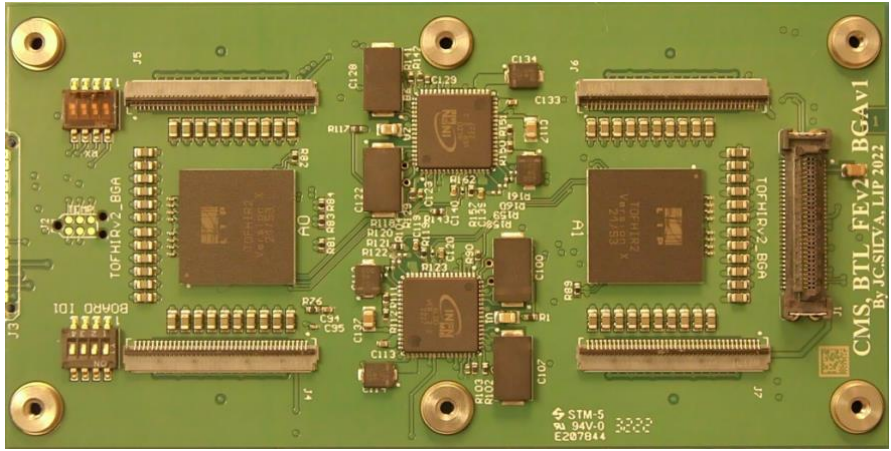

**Figure 10.** The BTL front-end board prototype with two TOFHIR2 and two ALDO2 ASICs.

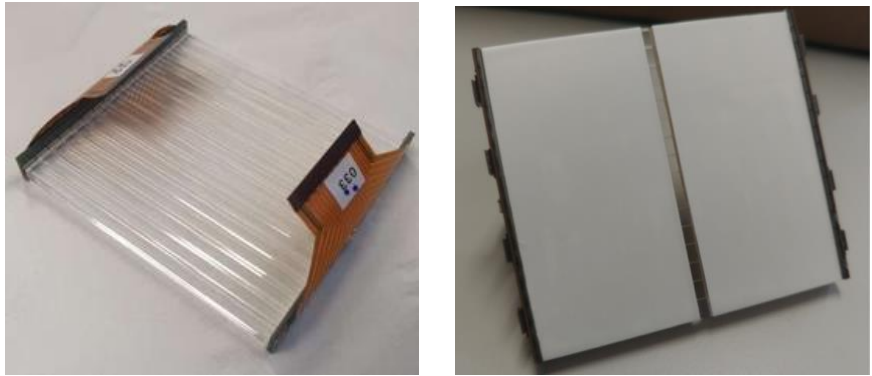

**Figure 11.** The BTL sensor module composed of 16 LYSO crystal bars glued at both ends to linear arrays of 16 SiPMs (left). The slit on the external wrapping allows shining the UV laser directly on the crystal bars to induce photo-excitation in the LYSO (right).

Measurements were performed using laser pulses generated by the HPK PLP-10 Picosecond Light Pulser (wavelength 405 nm) or by the NKT Photonics PIL1-037-40 UV laser (wavelength 375 nm). The laser power is adjusted to obtain a given number of photoelectrons per pulse. The calibration of the pulse amplitude is obtained by measuring the SiPM output current. The tests of EoO performance are made with un-irradiated SiPMs. DarkCount noise is emulated by using a blue LED. Measurements with blue laser beam impinging directly on the SiPMs were also performed. Figure 12 shows a picture of a typical test setup inside a thermally controlled black box used in the measurements.

Access to data via the ASIC I/O digital links is done using the PETsys Readout System [44]. Flexible cables connect the test board to the FPGA in the readout motherboard. The readout system allows the readout of two TOFHIR2 ASICs. The system provides the front-end board with all the necessary power, SiPM bias voltages, configuration, and readout. The TOFHIR2 supply voltage and the SiPM bias voltage are regulated by the ALDO2 ASIC [45].

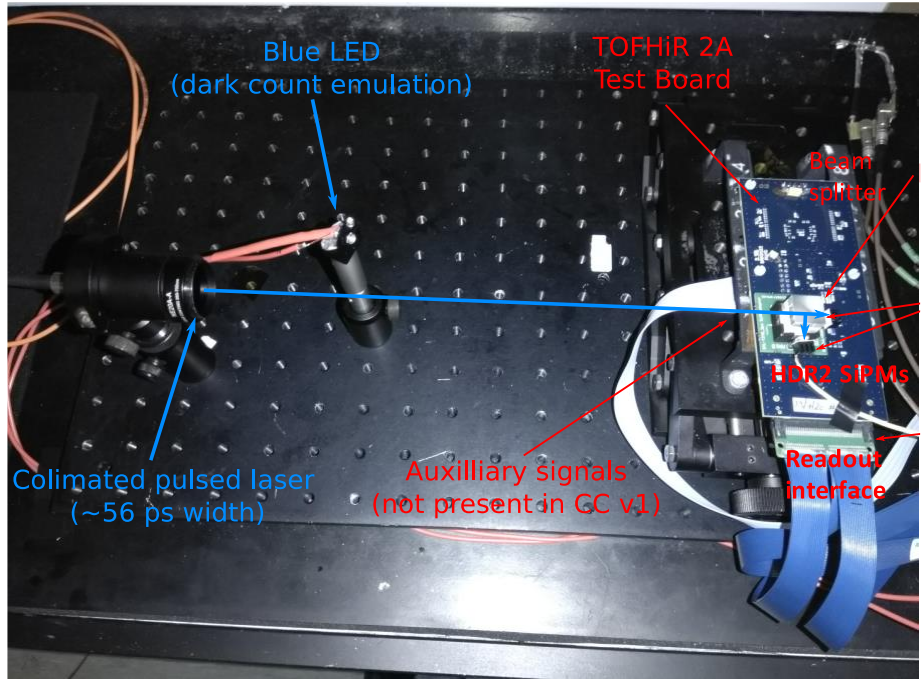

**Figure 12.** Example of the test setup used in the TOFHIR2 characterization measurements. In this case, we have laser (signal) and blue LED (DarkCount noise) light impinging directly on SiPMs mounted on the TOFHIR2A test board. A beam splitter is used to produce synchronous signals.

## 6. Characterization results

### 6.1 Power consumption

The measured consumption of the 32-channel TOFHIR2C ASIC is 455 mA when the chip is powered, the clock is active, the 32 channels are firing at 2.5 MHz with the digital test pulse and 3% of the events are accepted by the L1 trigger and transmitted to the readout system.

In this measurement there is no dynamic consumption on the analog frontend and there are no low energy hits activating the TDC/QDC and respective digital logic. From simulation, the corresponding consumption in BoO conditions is about 35 mA. The analog dynamic contribution is not expected to change significantly along the detector lifetime since the decrease in signal yield is compensated by the increase of DCR. Therefore, taking this into account, we estimate that the TOFHIR2C consumption in operating conditions is 490 mA.

In TID radiation tests we have observed an increase of current consumption of the order of 5-10%, but we have shown that the natural annealing due to pauses of the accelerator bring the radiation effects to below 1% (see section 7.1). Therefore, we neglect the radiation effect in the power consumption.

In conclusion, as the supply voltage is 1.2 V and the chip has 32 channels, the estimated power consumption of TOFHIR2C per channel (static + dynamic) is 18.4 mW.

### 6.2 Pulse shape and noise

We measure the pulse shape tuning the UV laser intensity to yield a number of photoelectrons representative of BoO conditions. The shape of the pulses is obtained by scanning the discriminator threshold (LSB of 1.25 µA). The time of the leading and trailing edges of

– 16 –

discriminator output pulse are measured by the TDC1 and TDC2, respectively, allowing the reconstruction of the pulse shape. Using this method, the peak of the pulse is not observed due to the configurable range of the discriminator threshold. Figure 13 shows a good agreement between simulation and data for the rising edge of the pulse. The slew rate in the rising edge at the level of the optimum threshold for timing measurement is 28.6 µA/ns (to be compared to 28.0 µA/ns in simulation). The observed small discrepancy in the pulse trailing edge is ascribed to a different DLED delay and trim (see section 3.1) in the chip used in the measurement and in the simulation.

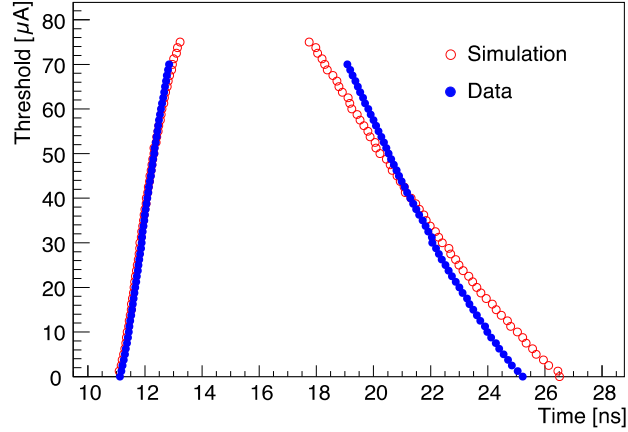

**Figure 13.** Pulse shape at the input of the discriminator reconstructed with a threshold scan as obtained with simulation data and experimental data using TOFHIR2X.

The combined contributions of the electronics noise and TDC to the time resolution are estimated with blue laser light shining directly on two naked SiPMs using a beam splitter. The coincidence time resolution (CTR) is measured between the two channels and the channel time resolution is obtained as CTR/$\sqrt{2}$. Figure 14 shows the measured and the simulated channel time resolution as a function of the pulse slew rate ($dI/dt$). Measurements and simulation are in good agreement. A fit of the data points with the function:

$$\sigma_t = \sigma_{noise}/(dI/dt) \oplus \sigma_{TDC}$$

yields $\sigma_{noise}$= 0.36±0.03 µA and $\sigma_{TDC}$= 12±1 ps. The noise measurement agrees with the simulation estimation of the electronics noise at the input of the discriminator (0.42 µA). The TDC resolution agrees with the direct measurement presented in the next section.

The results reported in this section were obtained with TOFHIR2X and SiPMs of cell size 15×15 µm².

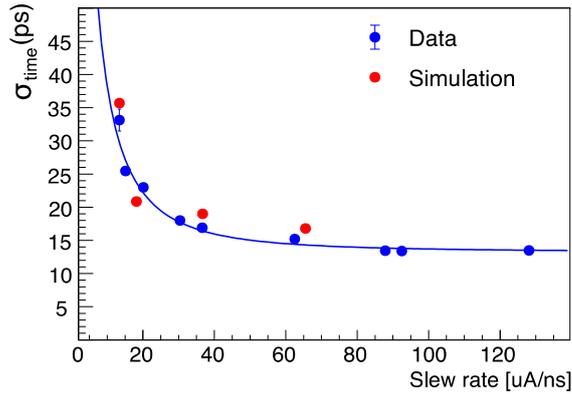

**Figure 14.** Time resolution of laser pulses directly detected by a SiPM as a function of the slew rate of the pulse rising edge in data and in simulation. The blue line represents a fit of the data points (see text). Results obtained with TOFHIR2X.

### 6.3 TDC performance

The TDC calibration uses an external digital pulse synchronous to the system clock and distributed internally to all TDC inputs in the chip. The time of the test pulse relative to the clock edge, defined in the FPGA of the readout board, is scanned in steps of ten picoseconds. The distribution of the TDC binning derived from the calibration data is shown for 2032 TACs in 4 different ASICs in Figure 15. The average bin size is 11.3 ps which matches well the value of 10 ps expected from simulation. It is worth noting the low dispersion of TDC binning (r.m.s. = 0.4 ps).

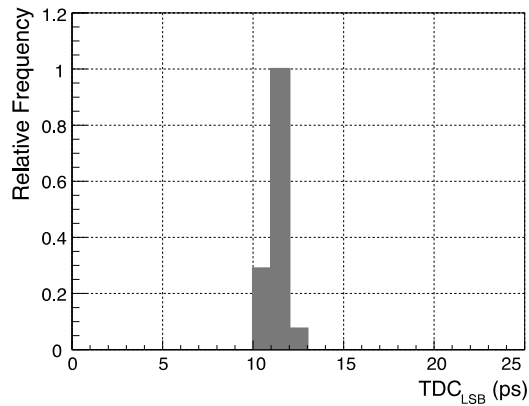

**Figure 15.** Normalized distribution of the TDC bin measured for 2032 TACs in 4 different ASICs.

The linearity of the TDC is derived from the code density distribution measured with random digital pulses following a uniform distribution in time. Figure 16 shows the Differential Non-Linearity (DNL) and the Integral Non-Linearity (INL) as a function of the TDC code obtained for two different TACs. Before linearity corrections, DNL is less than ± 0.5 LSB and the INL is less than ±2 LSB.

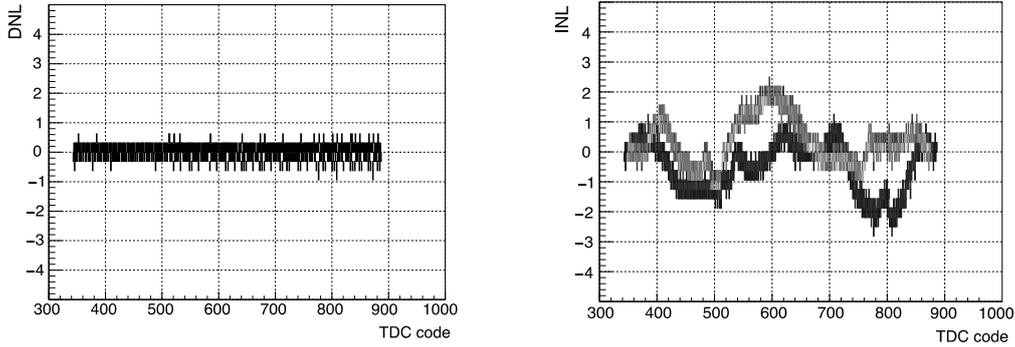

**Figure 16.** DNL and INL as a function of the TDC code obtained for two different TACs.

The TDC resolution is derived from coincidences between two TDCs in the chip receiving a common test pulse. The average coincidence time resolution (CTR) estimated with 28 TDC pairs is 18.8 ps, from which we derive a TDC resolution of 13.3 ps as shown in Figure 17. In this measurement, the dispersion of the resolution is 5.3% r.m.s.

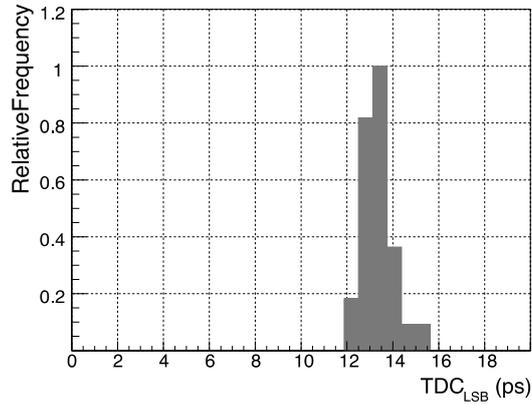

**Figure 17.** Normalized distribution of TDC resolution.

### 6.4 Time resolution of MIP equivalent pulses

In BTL, the timing of a MIP particle is obtained from the average of the two measurements in a single LYSO bar. The bar time resolution may be derived from the CTR of the two channels in the crystal bar ($\sigma_{bar} = CTR/2$)[13].

We performed measurements with the UV laser tuned to generate a LYSO pulse with 9800 photoelectrons when the SiPMs are operated at overvoltage of 3.1 V (SiPM gain $9.7 \times 10^5$), characteristic of BoO conditions. Figure 18 shows the bar time resolution as a function of the discriminator threshold for the optimal settings of the delay line in the noise suppression circuit. At the optimal threshold, we obtain a resolution of 24 ps, compatible with the simulation results.

---

[13] Equivalent to the average of two measurements for a fixed impact point position along the bar if there are no correlated uncertainties between the two ends.

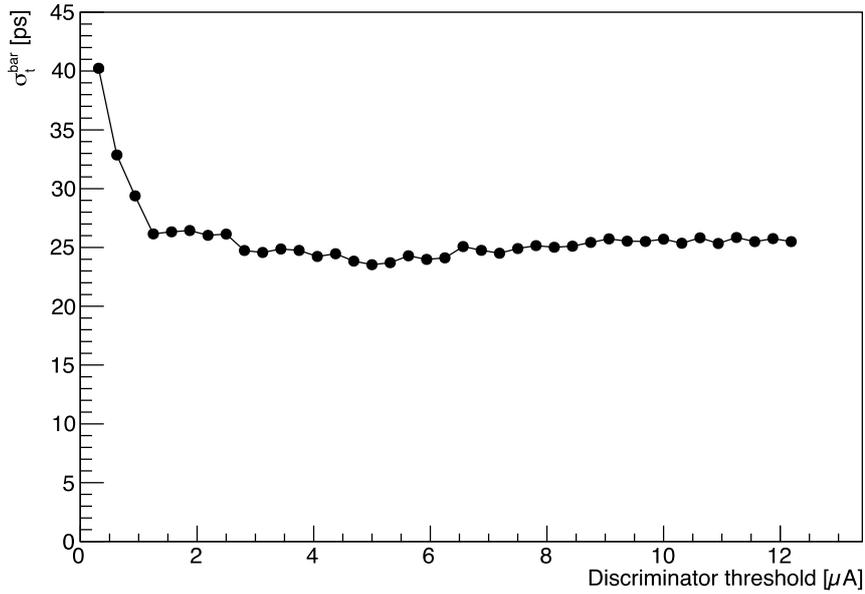

**Figure 18.** Time resolution of LYSO pulses characteristic of BoO as a function of the discriminator threshold.

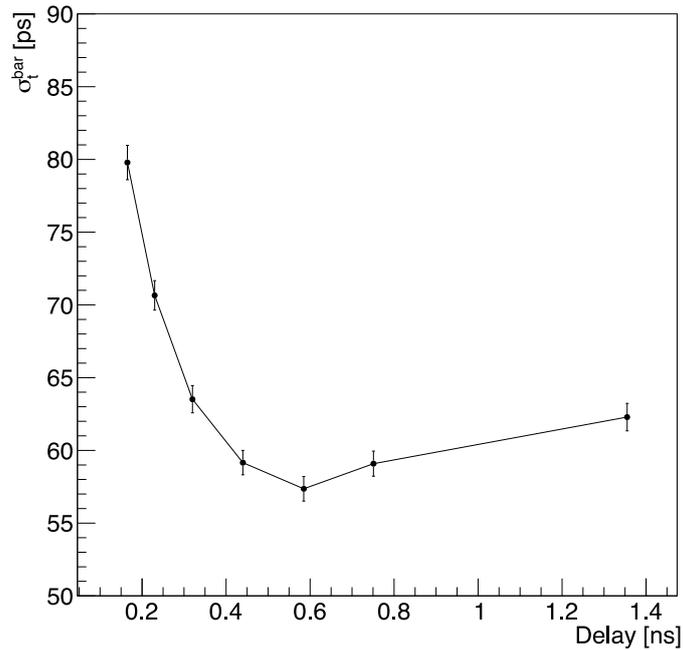

**Figure 19.** Time resolution of LYSO pulses characteristic of EoO as a function of the delay line in the DarkCount noise suppression circuit (right).

To reproduce the large DCR at EoO, we use background LED blue light emulating the SiPM DarkCount noise. The calibration of the LED light is established by measuring the SiPM current as a function of the LED voltage. The SiPM current is converted to equivalent DCR by considering the SiPM gain at the operating overvoltage. We measure the channel time resolution of laser pulses characteristic of EoO conditions, while illuminating the SiPMs with blue light emitted by the LED. Figure 19 shows the time resolution as a function of the delay in the DarkCount noise suppression circuit for pulses with 4400 p.e. and SiPM gain of $3.6\times10^5$ and equivalent DCR of 18 GHz. In this measurement, the SiPM is operated at an overvoltage of 1.0 V as foreseen at EoO to limit the DarkCount noise. For the optimal setting of the delay line (590 ps), we obtain a resolution of 58 ps. The signal yield and SiPM parameters used in the measurements reported above are representative of measurements performed with the revised sensor module, assembled both with non-irradiated and irradiated SiPMs, in a test beam at CERN[14].

In order to assess the impact of the channel rate on the time resolution, we performed measurements of the time resolution in a single channel (one side of the LYSO bar) as a function of the average rate of UV laser pulses. The measurement was done in a single channel to reach 2.5 MHz without saturating the chip output bandwidth. The laser is triggered with a pseudo-random sequence of pulses. We chose an amplitude of the laser pulses such that the bar resolution is typical of EoO. We have observed that the laser is not able to generate pulses of constant amplitude when separated by less than 250 ns, therefore we removed from the analysis these pulses. Figure 20 shows that the time resolution remains stable as a function of the effective average rate of pulses used in the study.

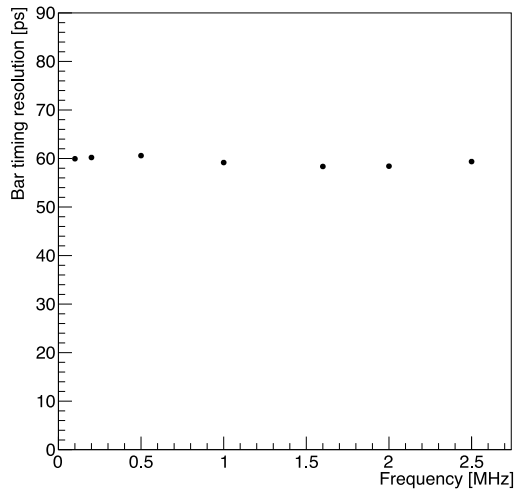

**Figure 20.** Time resolution as a function of the laser pulse rate (see text).

## 6.5 Energy of MIP equivalent pulses

We have performed measurements of the event energy as given by the TOFHIR2 pulse charge integration using UV laser pulses impinging in the sensor module. The laser power was tuned to deliver a given number of photoelectrons per pulse as derived from the measurement of the SiPM current. We performed measurements at two values of the SiPM overvoltage, namely 3.1 V and 1.0 V, typical of BoO and EoO operations. In each of the two cases, we configured the QDC attenuator to match the expected range of pulse amplitude.

---

[14] Test beam results will be described in a future publication.

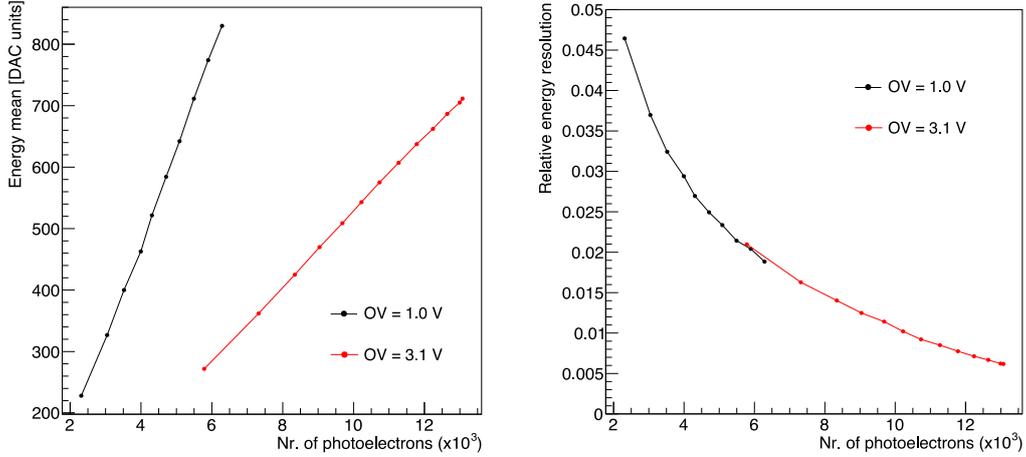

**Figure 21.** Average energy (left) and relative energy resolution (right) as a function of the number of photoelectrons in the signal for two values of the SiPM overvoltage, typical of BoO (3.1V) and EoO (1.0V) conditions, respectively.

In these measurements, as shown in Figure 21 (left), we observe a linear response in the amplitude range typical of BoO (overvoltage 3.1 V) and EoO (overvoltage 1.0 V). We also observe in Figure 21 (right) that in all cases the relative energy resolution is below the specified 5%.

**6.6 Rate performance**

The rate performance of the TOFHIR2 chip was assessed with simulations and with test pulse measurements.

The eight-fold analog buffering in TACs and QAC before digitization is potentially a source of inefficiency due to buffering overflow. We simulated pseudo-random pulses at the channel input of two types: MIP pulses at an average rate of 2.5 MHz, which are digitized by the ADC, and low energy pulses that are buffered in the TAC but rejected for digitization. The rate of the low energy pulses was varied between 5 MHz, the estimated rate of energy deposits above 100 keV in BTL, and 32 MHz, the average bunch crossing rate at LHC. In the simulation, we conservatively assumed 100 ns per measurement digitization, and considered the cases of two measurements (one timestamp and one charge) and three measurements (two timestamps and one charge). As shown in Figure 22, the channel inefficiency is 1% at the nominal rate of low energy events of 5 MHz raising to 4% at the rate of 32 MHz. Configuring the chip to perform only two measurements, the inefficiency at 5 MHz becomes negligible and the maximum inefficiency is below 3%.

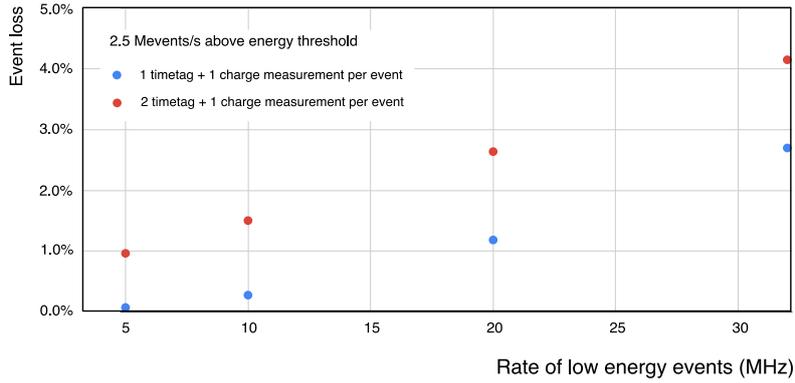

**Figure 22.** Channel inefficiency for a MIP rate of 2.5 MHz as a function of the rate of low energy events.

Using the TOFHIR2 test system with adequate firmware, we could inject pseudo-random test pulse sequences in the ASIC that emulate BTL operation parameters, namely average 2.5 MHz digitized events per channel and 3% L1 trigger acceptance (2.67 MHz events transmitted per ASIC). Every 25 ns a pseudo random number, x, is generated. If x < tp a test pulse is generated triggering the TDC, and if x < tgr a L1 trigger acceptance is generated, where tp and tgr are tunable parameters. The L1 trigger acceptance is transmitted with a 12.125 µs delay, the expected latency in the experiment. L0 is not used in these measurements. Correct ASIC L1 latency setting can be identified by sweeping the setting and observing event counts (Figure 23).

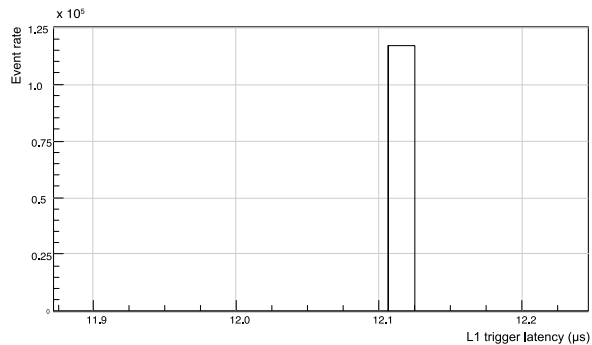

**Figure 23.** Event output rate as a function of the L1 latency setting.

When generating test events in a single channel, we could observe that the rate of transmitted events follows the trigger acceptance rate up to the maximum chip output rate, as shown in Figure **24**. Generating test events simultaneously in various channels we have observed that the output event rate grows linearly with the number of activated channels up to the maximum output rate of 2.67 M events/s (Figure 25).

These results prove that TOFHIR2 is capable to operate in the high-rate conditions of BTL at the HL-LHC. Indeed, a GEANT4 detector simulation with 200 pile-up events per crossing has shown that the maximum channel occupancy for energy deposits above 1 MeV is 8% [2], corresponding to an average of 2.56 activated channels per crossing. Therefore, a L1 trigger rate of 750 kHz as planned in CMS leads to an output rate of 1.92 M events/s, which is 72% of the chip output bandwidth (one Tx link).

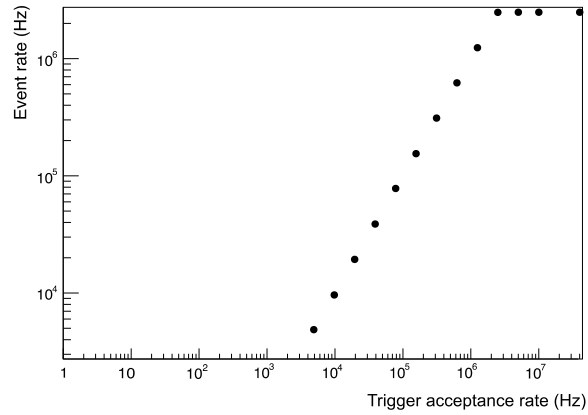

**Figure 24.** Channel event rate as a function of the L1 trigger rate.

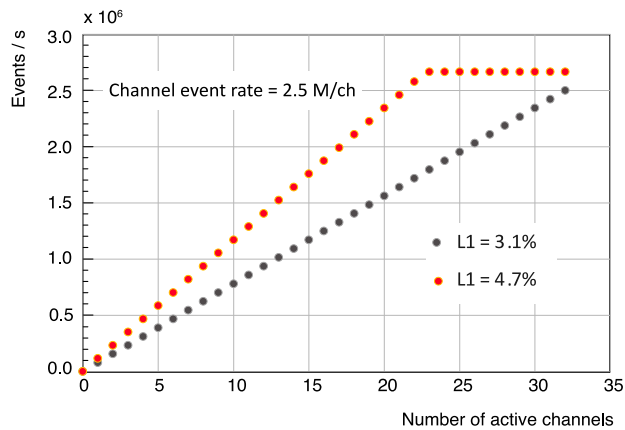

**Figure 25.** Output event rate as a function of the number of activated channels for an input rate of 2.5 MHz per channel and for two values of the L1 trigger acceptance.

## 7. Radiation tests

### 7.1 TID radiation tests

The resistance to TID radiation of the three TOFHIR2 prototypes was tested at the X-ray irradiation facility at CERN (Figure 26-left). The maximum expected dose in the barrel MTD is 2.9 Mrad. At each campaign, we have irradiated two ASICs, the first one up to 7 Mrad in steps of 0.5 Mrad and another up to 3 Mrad in steps of 1 Mrad. Several measurements were performed between steps, as well as after the irradiation. The irradiations were performed at -25 ºC.

In the first campaign, we irradiated TOFHIR2A fabricated in foundry B. Using the test pads, we measured the current consumption and the bandgap voltage, and we performed several DAC voltage scans (amplifier baseline, discriminator threshold, QAC/TAC baseline, ALDO DAC). Using the TOFHIR2 data acquisition, we monitored the front-end noise from a scan of the T1 discriminator threshold, measured the shape of the internal analog test pulse using a scan of the T2 discriminator threshold, and performed the TDC and QDC calibrations in all TACs and QACs in the two ASICs. The time interval between steps was of the order of 10 minutes to prevent annealing effects. The same measurements were performed 12 hours after the irradiation in order to assess the effect of annealing at room temperature. Additionally, the time resolution with laser pulses was measured using the ASIC irradiated at 3 Mrad.

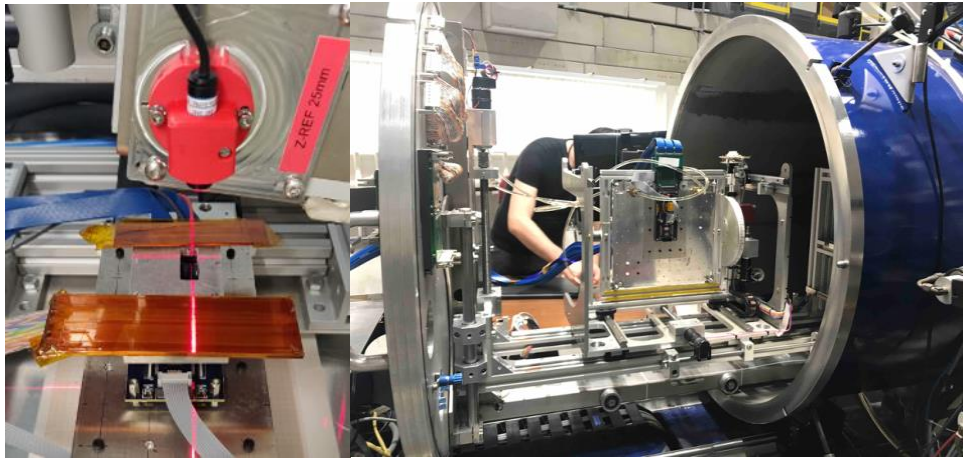

**Figure 26.** Setup used in TID radiation tests at the X-ray irradiation facility at CERN (left). Set used in SEE radiation tests at the Heavy Ion Facility (HIF) in Louvain-la-Neuve (right).

We observed effects due to the large leakage current at doses of ~1 Mrad reported for the technology CMOS 130 nm produced in foundry B [33]. An increase of 20% of the current consumption after a dose of 1 Mrad followed by a decrease for larger doses was seen. In agreement with previous observations [33], the current consumption returns to the original value after 12 hours annealing at room temperature. We also observed that the increase of leakage in NMOS transistors originates a decrease by 20% of the DACs voltage range at 1 Mrad.

On the other hand, we observed negligible or minor effects on the frontend amplifiers, TDC and QDC up to 7 Mrad. Figure 27 shows the TDC time resolution as a function of the irradiation dose and Figure 28 shows the time resolution obtained with SiPMs and blue laser pulses as a function of discriminator threshold before and after irradiation at 3 Mrad. Within measurement uncertainties no sizable effects are observed.

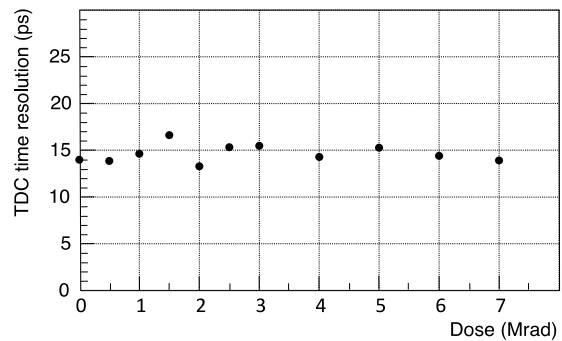

**Figure 27.** TDC time resolution as a function of TID measured with TOFHIR2A.

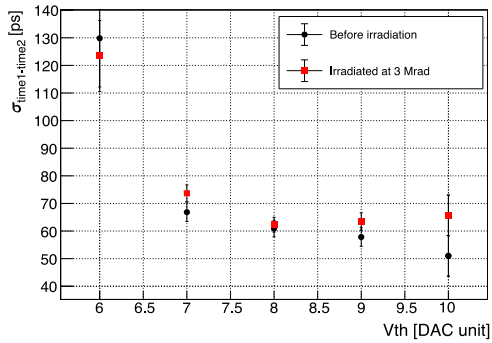

**Figure 28.** Coincidence time resolution obtained with naked SiPMs and blue laser pulses with amplitude 40 mV as a function of discriminator threshold before and after irradiation at 3 Mrad measured with TOFHIR2A.

While TOFHIR2A front-end is not the current-mode front-end of the final chip, and the foundry used for its fabrication is the one yielding the largest leakage current, the above results gave confidence that the techniques used in the TOFHIR2 design to mitigate TID radiation effects were effective. The same techniques have then been used in the other versions of the chip.

The version TOFHIR2X, implementing the final analog front-end design, was fabricated at foundry A, the same where the final chip production was made. We expected the effects due to transistor leakage to be smaller, since the increase of leakage current at 1 Mrad in foundry A is considerably smaller than in foundry B. A similar irradiation campaign was carried out by irradiating the ASIC up to 7 Mrad in steps of 0.5 and 1.0 Mrad. Before irradiation (at 26 ℃), during irradiation at each step, and 15 hours after irradiation (annealing at room temperature), the same measurements as done with TOFHIR2A were performed.

As for TOFHIR2A, we observe variations due to the leakage current, however of smaller amplitude. For example, in TOFHIR2X the range of the ALDO DAC decreases by 5%, compared to 20% in TOFHIR2A, and almost fully recovers after annealing at room temperature (Figure 29 left), while the design did not change between the two versions.

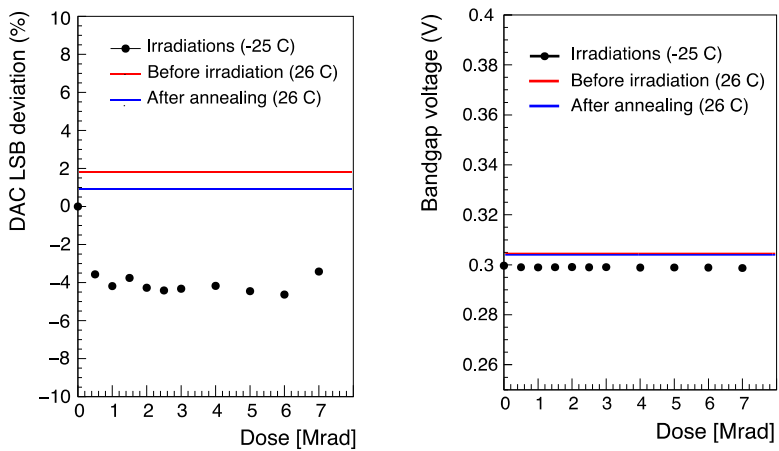

**Figure 29.** Relative deviation of the ALDO DAC LSB as a function of TID measured with TOFHIR2X (left). Bandgap voltage as a function of TID (TOFHIR2X) (right).

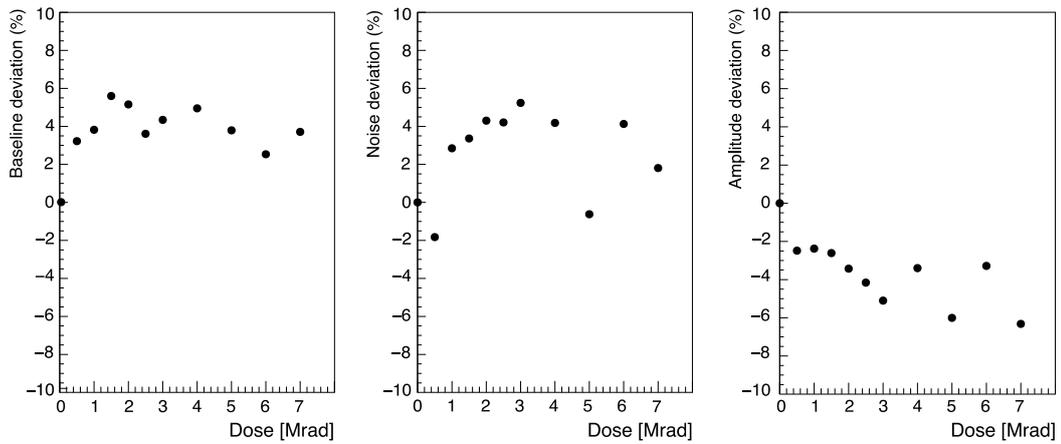

**Figure 30.** Deviation of the front-end baseline, noise and test pulse amplitude relative to the measurement at -25°C before irradiation, as a function of TID.

On the other hand, we observe negligible or minor effects on the frontend amplifiers, TDC and QDC up to 7 Mrad. Figure 29-right shows the that the bandgap voltage used as reference in the whole chip is stable up to 7 Mrad. Figure 30 shows the dependence with radiation of the current-mode baseline and noise where we observe variations of 4-5%. It also shows a reduction of the test pulse amplitude of ~6% after 7 Mrad, as it was already observed with TOFHIR2A.

A campaign of measurements performed with different doses and time intervals between measurements emulating the real annealing scenarios resulting from the normal pauses during the accelerator operations, has shown that no effects larger than 1% remain. Within measurement uncertainties no other sizable effects are observed. In particular, as shown in Figure 31, the time resolution of the internal analog test pulse remains stable up to 7 Mrad. In this measurement, the pulse amplitude was tuned to get a resolution of the order of 50 ps from electronics noise, typical of EoO conditions.

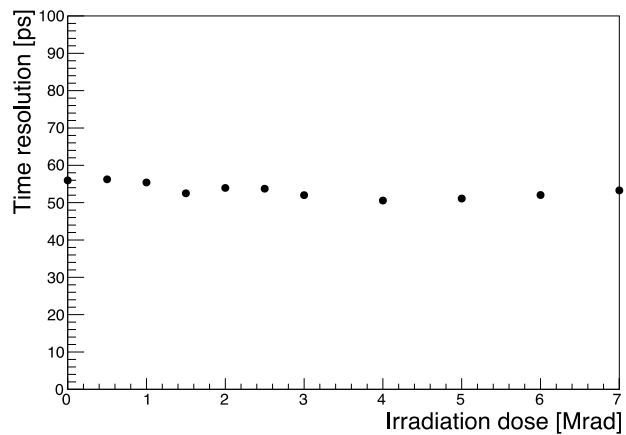

**Figure 31.** Time resolution of the analog test pulse as a function of TID.

## 7.2 SEE radiation tests

To protect against Single Event Effects (SEEs), TOFHIR2 implements Triple-Mode Redundancy (TMR), which triplicates flip-flops (FF) and implements majority-voting in the configuration bits (15'558 flip-flops), readout logic and clock/reset trees. FF upsets are identified and corrected by the majority logic. The total number of corrected errors are counted in the chip. Uncorrected errors could arise when a single particle changes the state of two out of three TMR flip-flops. Uncorrected errors are identified by reading back the configuration. Possible single event transients (SET) in the clock and resync tree could originate spurious chip resets or synchronization errors.

The test of SEEs due to localized ionization in TOFHIR2 chips was performed at the Heavy Ion Facility (HIF) in Louvain-la-Neuve (Figure 26-right). We performed irradiations with different ion beams ($^{53}Cr^{16+}$, $^{36}Ar^{11+}$, $^{27}Al^{18+}$, $^{22}Ne^{7+}$, $^{13}C^{4+}$, $^{84}Kr^{25+}$) and different incident angles (0°, 30°, 45°) covering a wide range of linear energy transfer (LET) that allow the SEU errors to be extrapolated to the LHC environment. The maximum beam flux used was $1.4 \times 10^4$ ions/s/cm$^2$ with $^{53}Cr$ ions. The beam homogeneity of 10% on a diameter of 25 mm ensured uniform irradiation of the whole chip area. At the various steps of the chip development, we performed HI irradiation campaigns, namely with TOFHIR2A, TOFHIR2B and TOFHIR2C with consistent results. The results presented below were obtained with the final version of the chip.

During the HI irradiations, we operated TOFHIR2 with the readout system described in section 5 and acquired events generated by external test pulses. During testing we monitored three ASICs, the ASIC being irradiated, the other ASIC in the test board, and an ASIC outside the beam vacuum chamber. The same test pulses are distributed to all ASICs such that the output data generated should be identical. As additional debug features, we integrated logic analyzers on the ASIC receiver logic in the readout firmware and have implemented firmware capable of recording "garbage data", if any, for post-analysis.

We checked the event data content, particularly the value of the clock counter (coarse time tag), to identify single event transients that may create glitches in the clock or synchronization signals. Additionally, every five seconds, we read the SEU counter and all configuration bits. These observations allowed to monitor the cumulative SEU counter (corrected SEU errors), the number of configuration bits flipped (non-corrected errors) and the number of errors in the coarse time tag (synchronization errors).

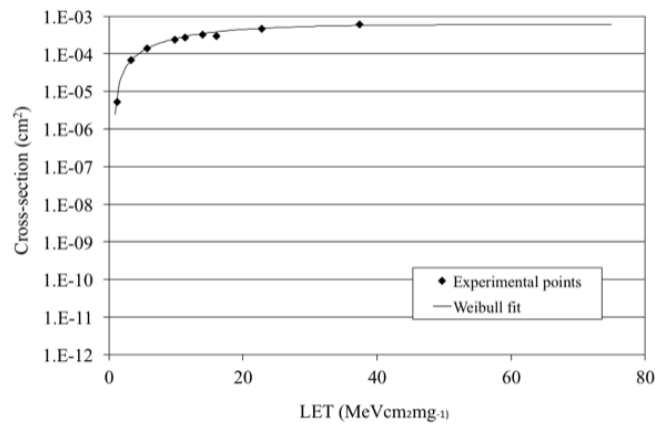

**Figure 32.** Measured cross-section of corrected SEU errors as a function of LET and the Weibull distribution fit (see text).

Figure 32 shows the measured cross-section of corrected SEU errors as a function of LET. The cross-section, given by the number of errors divided by the beam fluence, is parametrized by a Weibull distribution:

$$\sigma = \sigma_0 \cdot \left(1 - e^{-\left(\frac{L-L_0}{w}\right)^s}\right)$$

with the following parameters: $\sigma_0 = 6.0\times10^{-4}$ cm$^2$, $L_0 = 8.0\times10^{-1}$ MeV·cm$^2$·mg$^{-1}$, $w$=15 and $s$=1.2. This cross-section is used as input to a simulation model of the HL-LHC environment. The resulting cross-section at HL-LHC is about $4\times10^{-10}$ cm$^2$/chip, indicating a factor of $10^6$ between HI and HL-LHC cross-sections for the technology used.

We have observed a few (<5) uncorrected configuration bit errors and synchronization losses in runs with highly charged ions. The ratio of uncorrected/corrected errors is of the order of 1/1000, therefore the cross-section for uncorrected configuration bits is around $4\times10^{-13}$ cm$^2$/chip. We measured a similar cross-section for synchronization losses. The fact that the errors were observed with highly charged ions indicates that the above cross-section is a pessimistic estimation. The average MIP flux in the BTL detector is $3.3\times10^5$ cm$^{-2}$s$^{-1}$, therefore the estimated rate of uncorrected errors or synchronization losses is $1.3\times10^{-7}$ error/chip/s.

In a few cases we observed data transmission lockup in runs with heavy ions, likely due to uncorrected SEUs in the logic controlling the readout FIFOs. To quantify this occurrence, an irradiation was done in blocks of 5 s with data being acquired in the last 0.5 s. The digital test pulse is sent every 1024 clock cycles (6.4 μs). A lockup is defined as not getting any events from the ASIC under test in the last 1000 frames (6.4 ms) of a 0.5 s acquisition. A reset is applied at the end of each block.

In the runs with $^{53}$Cr beam, for a total fluence of $3.7\times10^6$ part/cm$^2$, we observed 23 data transmission lockups, from which we estimate a lockup cross-section of $6.2\times10^{-12}$ cm$^2$/chip at HL-LHC. This would translate into a lockup rate of $2\times10^{-6}$ lockup/chip/s. Data taken with the heavier ion $^{84}$Kr showed a lockup cross-section 1.8 times higher, as expected. The recoil ions in silicon have considerably lower charge, therefore the indicated rate is an upper limit.

The rate of configuration corruption, loss of synchronization and lockup of data transmission as reported above is small but not completely negligible. The operation of the BTL detector will require periodic configuration reloading and resynchronization, to be done typically every hour. In TOFHIR2, data itself is not TMR protected. Based on the measurements performed with heavy ion beams, we estimate the fraction of event data lost or corrupted to be of the order of $10^{-9}$.

## 8. Conclusions

The TOFHIR2 readout chip for the CMS barrel MIP Timing Detector was developed. The measurements of time resolution with TOFHIR2 and BTL sensor modules performed in laboratory test setups match very well the simulation expectations. In particular, the time resolutions, obtained for BoO (24 ps) and under stringent EoO conditions characterized by very large DCR and smaller pulses (58 ps) fulfill the BTL requirements. Successful TID and SEE radiation tests were also performed. This critical block of the CMS timing detector was proved to be feasible and to achieve the required performance.

**Acknowledgments**

We thank the MTD sensor team for providing the SiPMs and sensor modules used in the measurements and more generally the MTD project for all the support provided to the TOFHIR2 development. We also thank the CMS collaboration for creating the conditions that made this work possible.

We thank the Microelectronics Group at CERN for all the support and advice in the development of TOFHIR2, as well as for making available the radiation tolerant bandgap block and the TMRG tool.

We thank Federico Faccio, Giulio Borghello and André David for helping in the preparation of the radiation tests. In particular we are thankful to Federico Faccio for the extrapolation of the heavy-ion SEE data to the LHC environment presented in this paper.

We also thank the coordinators of the CERN X-ray irradiation facility and of the Heavy Ion Facility (HIF) in Louvain-la-Neuve for making available these facilities and for the help provided.

AGH has received funding from the Polish Ministry of Science and Higher Education under contract No 5179/H2020/2021/2, and from the European Union's Horizon 2020 Research and Innovation programme under grant agreement No 101004761 (AIDAinnova).